\documentstyle[12pt,aasms4]{article}

\def\reference{\par\noindent\hangindent=1cm\hangafter=1}

\newcommand{\eq}{\begin{equation}}
\newcommand{\ee}{\end{equation}}

\def\t0{\theta_{\circ}}

\def\be{\begin{equation}}
\def\en{\end{equation}}

\def\gapp{\ \lower 3pt\hbox{${\buildrel > \over \sim}$}\ }
\def\lapp{\ \lower 3pt\hbox{${\buildrel < \over \sim}$}\ }

\lefthead{Chen \& Jiang}
\righthead{Outer Solar System}

\begin{document}

\title{The Size Distribution of the Kuiper Belt Objects}

\author{Cheng-Pin Chen \& Ing-Guey Jiang}

\affil{Academia Sinica, Institute of Astronomy and Astrophysics, Taipei, 
Taiwan}

\authoremail{jiang@asiaa.sinica.edu.tw}

\begin{abstract}

The size distribution is determined directly from the current known
Kuiper Belt Objects (KBOs). We found that there is a peak for the size 
distribution around 230 km in diameter. For the objects larger
than 230 km,  the KBOs populate as $N(s) 
\propto  s^{-4}$ 
as we see in Jewitt et al. (1998). For the objects smaller than 
230 km, we found $N(s) \propto s^{3}$. This result from the current
observational data is new and very different from the conventional
concept of the KBO size distribution. Though the true size distribution
might change from what we have after more KBOs be discovered,  
we argue that the existence of a peak around 150 km to 250 km is likely.  
 
\end{abstract}


\section{Introduction}

After the first KBO was discovered in 1992 (Jewitt \& Luu 1993),
it is known that the outer solar system beyond Neptune is populated by 
small bodies. The total number of KBOs is increasing quickly and the general 
picture of the outer solar system is changed. There are three classes of KBOs
according to the current observations. One of the population are the
Resonant KBOs. They are trapped in 3:2 mean-motion resonance with Neptune
at 39.4 AU. About one-third of the total population are in this class.
The second class mainly occupies the region between 41 AU and 46 AU with small
eccentricities. They are called the Classical KBOs and about two-third of 
KBOs belong to this class. Finally, several KBOs which behave 
very differently from the above two were discovered since 1996 
(Luu et al. 1997)
and form a new class. They are Scattered KBOs and they are moving on
the large semi-major axises, highly eccentric, and inclined orbits.
These KBOs of new dynamical class could originate from the scattering
by Neptune.

One of the most important properties of KBOs is the size distribution  
because it tells the possible total number of KBOs and also gives the hint
of density profile of Solar nebula. To extract the size distribution
from the luminosity function, Jewitt et al. (1998) used Monte Carlo models
to simulate the survey under the necessary assumptions about the 
geometric albedos and distance distributions of the KBOs. In principle,
they assume a power-law size distribution as $\propto r^{-q}$ and determine
the best value of $q$ by reproducing  the luminosity function from 
the  models. They found $q = 4.0 \pm 0.5$ and claimed this value of $q$
is consistent with $q \approx 3.5$ from what was expected from a
collisional production function in Dohnanyi (1969). Kenyon \& Luu (1999)
did some planetesimal accretion calculations in a single annulus at 35 AU.
They found that all models produce two power-law size distributions,
$q = 2.5 $ for radii $\lesssim 0.3-3$ km and $q=3$ for radii $\gtrsim
1-3$ km and the results are nearly independent of the initial mass in the
annulus.

Different from the approach in Jewitt et al. (1998), 
we first get the number of KBOs against
absolute magnitude histogram directly
from the data of current known KBOs. Then, we convert this 
histogram into the size distribution. In addition to the 
size distribution for all current known KBOs, the  
Plutinos and Classical KBOs' size distributions are studied
separately. These results would be in Section 2 and there are 
concluding remarks in Section 3.

\section{The Results}

We get the data of current known KBOs from the daily updated List of
Transneptunian Objects at the website:
http://cfa-www.harvard.edu/iau/lists/TNOs.html 
and Figure 1 is the plot of eccentricity against semi-major axis for all 
KBOs from these data. There are 282 KBOs in the list on 27th July 2000. 
We regard the Plutinos to be those objects with 
semi-major axises from 38.1 AU to 40.5 AU. 
The Scattering KBOs are those objects with eccentricities $e>0.3$
and semi-major axis $a> 40.5$ AU. All the rest are the Classical KBOs.
Therefore, 
63 of these current known KBOs
are Plutinos, 212 objects belong to Classical KBOs 
and 7 objects are Scattering KBOs.

We use the table of converting absolute magnitudes to diameters
at the same website to get the diameters. However, we transform
the values of diameters in the table to be the values for albedo = 0.04.
The histogram for number of KBOs against the absolute magnitude
is then ploted. This plot can be transformed to the 
histogram for number of KBOs against the diameters, i.e. the size 
distribution easily.

\subsection{All Known KBOs}

Figure 2 is the result for all current known KBOs
and it shows that there is a peak around 230 km. For those objects
larger than this peak value, the size distribution follows
\begin{equation}  
N(s) = \frac{150}{{(s/200)}^4}
\end{equation}
approximately, where $s$ is the diameters (km) of the objects.
 For those objects smaller than the peak value, the size
distribution becomes 
\begin{equation}  
N(s) = 90 {(s/200)}^3.
\end{equation}
It is encouraging that those objects larger than 230 km 
follow the same size distribution
as in Jewitt et al. (1998), i.e. $q=4$ power-law. However, the 
existence of this peak around 230 km and the different size distribution
for objects smaller than the peak value has never been mentioned and 
discussed before. One might argue that this current observational 
result is not complete enough to confirm this peak. The size 
distribution would be completely different after all 100 km size KBOs
are discovered. However, considering that all KBOs around or larger than 
100 km are all within the detection limit and thus the detection
probability is proportional to the number of objects.
Equation 1 gives us $N(100)/N(230) \approx 28$ and thus 
the current known 100 km size KBOs should be about 28 times more 
than 230 km size KBOs if all sizes of KBOs follow the size distribution
of Equation 1. It would be a puzzle that the current number
of 230 km size KBOs is larger than the number of 100 km size KBOs if the
size distribution is described by single power-law only.

Therefore, it is obviously that the peak we discover
from the data of current known KBOs should be real and Equation 2
might be a good approximation for objects between 70 km to 230 km, though
probably not for objects smaller than 70 km due to current 
detection limit.

\subsection{Plutinos}

As suggested by Malhotra (1995), the resonance sweeping due to 
the migration of Neptune might explain the trapping of KBOs 
into the 3:2 resonance. 
During this resonance sweeping process, 
it is possible that not all KBOs of different sizes 
can be trapped and become the Plutinos. 
Therefore, the size distribution 
for Plutinos might give us information about this.

According to the histogram in Figure 3,
we found that for the objects larger  
than 150 km, the size distribution can be fitted by 
a $q=4$ power-law 
\begin{equation}  
N_{\rm p}(s) = \frac{17}{{(s/200)}^4},
\end{equation}
for the objects smaller than 150 km, we have:
\begin{equation}  
N_{\rm p}(s) = 60 {(s/200)}^4.
\end{equation}
Comparing these results with Figure 2,
it seems that the distribution is narrower and sharper around the peak and 
the peak is about 150 km. This might imply
that there is a particular size, i.e. about 150 km, which is 
preferable to the resonance trapping process. This result
of Plutinos' size distribution might be important for the
dynamical calculations of resonance sweeping.

\subsection{Classical KBOs}

Different from the Resonant and Scattering KBOs, the Classical KBOs
might be the better tracer for the primordial Solar nebula because
there were not trapped or scattered by Neptune and thus closer to where
they were formed. Therefore, it would be interesting to know the size 
distribution for the Classical KBOs only.

According to the histogram in Figure 4,
we found that for the objects larger  
than 230 km, the size distribution can be fitted by 
\begin{equation}
N_{\rm c}(s) = \frac{140}{{ (s/200) }^4},
\end{equation}
for the objects smaller than 230 km, we have:
\begin{equation}
N_{\rm c}(s) = 65 {(s/200)}^3.
\end{equation}

These results are pretty much the same as the results for all KBOs
because at least two-third of all KBOs are in fact the Classical KBOs.

\section{Concluding Remarks}

We have used the current known data of KBOs 
to study their size distribution. We found that there is a peak
around 230 km and therefore the size distribution is unlikely 
to be described by one single power-law. What does this observational
result imply ?

The size distribution is related to the density of the Solar nebula
because during the formation stage of KBOs, 
the amount of available material determined the final size and therefore
the size distribution of KBOs. In fact,
the initial density of the Solar nebula is 
one of the most important factors to lead the evolution
of the Solar system.
One possibility is to use the current mass
distribution of planets to set up the 
`minimum mass Solar nebula' model, then rescale to a higher density value
but keep the same density profile
because there must be certain amount of mass loss during the planet
formation. From these assumptions, we can calculate that there
was more than 20 Earth mass between 30 to 50 AU. 
However, the total mass of discovered KBOs is only about 0.07 
Earth mass from the results in Section 2 
or up to 0.26 Earth mass by Jewitt et al. (1998). Because of this 
contradiction,
one might argue that 
the `minimum mass Solar nebula' model could be wrong, especially 
for the outer Solar system. 

One way to approach this problem is to study the formation of  
KBOs. Stern (1995) claimed that the timescale to produce the QB1s 
is longer than the disc lifetime and suggested that the present-day
disc is not representative of the ancient structure.
Kenyon \& Luu (1999) did some accretion calculations and conclude
that the `minimum mass solar nebula' would be enough to form large
KBOs in a reasonable timescale.
Thus, 
from these calculations of Stern (1995), Kenyon \& Luu (1999),
we do need the outer Solar system to be at least as massive as 
the `minimum mass Solar nebula' model to form the current observed
KBOs through the accretion. Therefore, the outer Solar system 
between 30 to 50 AU was at least two order of 
magnitude more massive and the depletion in this region 
definitely happened during the formation of the Solar system.

On the other hand, the accretion calculations in 
Kenyon \& Luu (1999) told us the size distribution for QB1s
can approximately described by single power-law when these
QB1s were formed. We therefore come to the conclusion 
that the size distribution should have changed during the depletion. 
 
Jewitt (1999) suggested that for collisions to have removed 99 percent
of the initial mass, one would have to postulate a steep initial
size distribution ($q \approx 5$ in which objects large enough to escape
commotion carry only 1 percent of the mass. 
This suggestion implies that the size distribution can be different
in the past and also that the larger objects can survive easier than the
smaller objects during the depletion. 
It is therefore possible that the peak we discovered was formed during this
depletion process. A complete simulation to combine the accretion process
and the orbital dynamics as in Jiang, Duncan \& Lin (2000) might help 
the understanding of this depletion and thus the evolution of size 
distribution.

\clearpage
\section*{Acknowledgments}

Ing-Guey Jiang wishes to take this opportunity to acknowledge the 
suggestions on Solar System Dynamics from Scott Tremaine and 
the discussion with David Jewitt, Eugene Chiang, Li-Chin Yeh, 
Doug Lin during
their visit at Academia Sinica, Institute of Astronomy and Astrophysics.
He also thanks the hospitality and discussion of Martin Duncan and
Doug Lin during his trips to Queen's University and UC Santa Cruz. 

Cheng-Pin Chen acknowledges
 the financial support of Undergraduate Summer Student
Program from Academia Sinica, Institute of Astronomy and Astrophysics.
\clearpage

\section*{REFERENCES}
\begin{reference}

\reference Dohnanyi, J. 1969, J. Geophys. Res., 74, 2531

\reference Jewitt, D. 1999, Annual Review of Earth and Planetary Science,
   27, 287
\reference Jewitt, D., Luu, J. X. 1993, Nature, 362, 730

\reference Jewitt, D., Luu, J. X. \& Trujillo, C. 1998, \aj, 115, 2125

\reference Jiang, I.-G., Duncan, M. J. \& Lin, D. N. C. 2000, in progress

\reference Kenyon, S. J. \& Luu, J. X. 1999, \aj, 118, 1101

\reference Luu, J. X., Marsden, B. G., Jewitt, D., Trujillo, C. A.,
Hergenrother, C. W., Chen, J., Offutt, W. B. 1997, Nature, 387, 573 

\reference Malhotra, R. 1995, \aj, 110,  420

\reference Stern, S. A. 1995, \aj, 110, 856

\end{reference}

\clearpage

\begin{figure}[tbhp]
\epsfysize 7.0in \epsffile{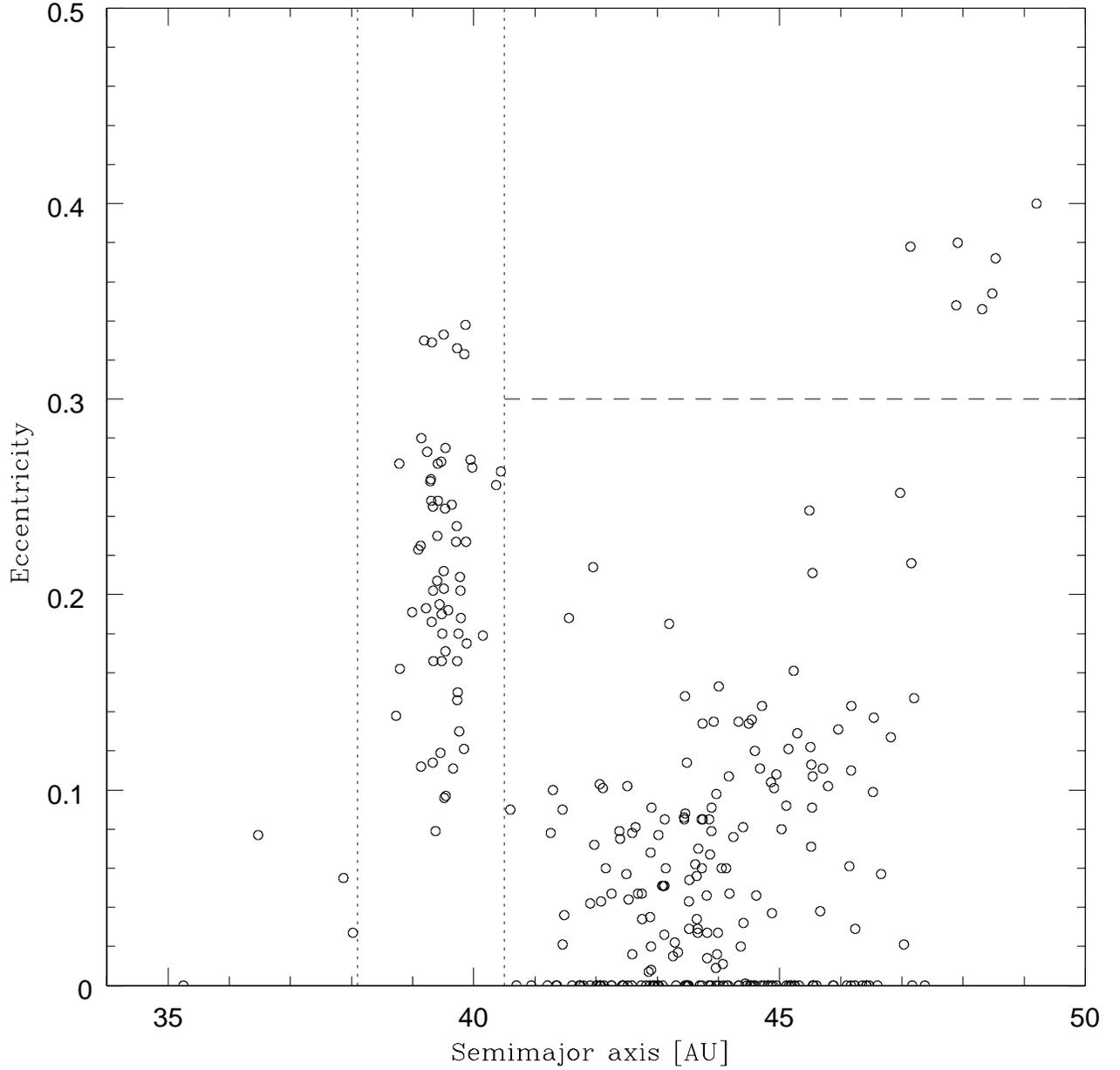}
\caption{The eccentricity against semi-major axis for all current
known KBOs. The Plutinos are those between two dotted lines, i.e.
with semi-major axises from 38.1 AU to 40.5 AU. The Scattering KBOs
are those with $e>0.3$ and semi-major axis $a> 40.5$ AU. All the rest 
are the Classical KBOs.}
\end{figure}

\begin{figure}[tbhp]
\epsfysize 7.0in \epsffile{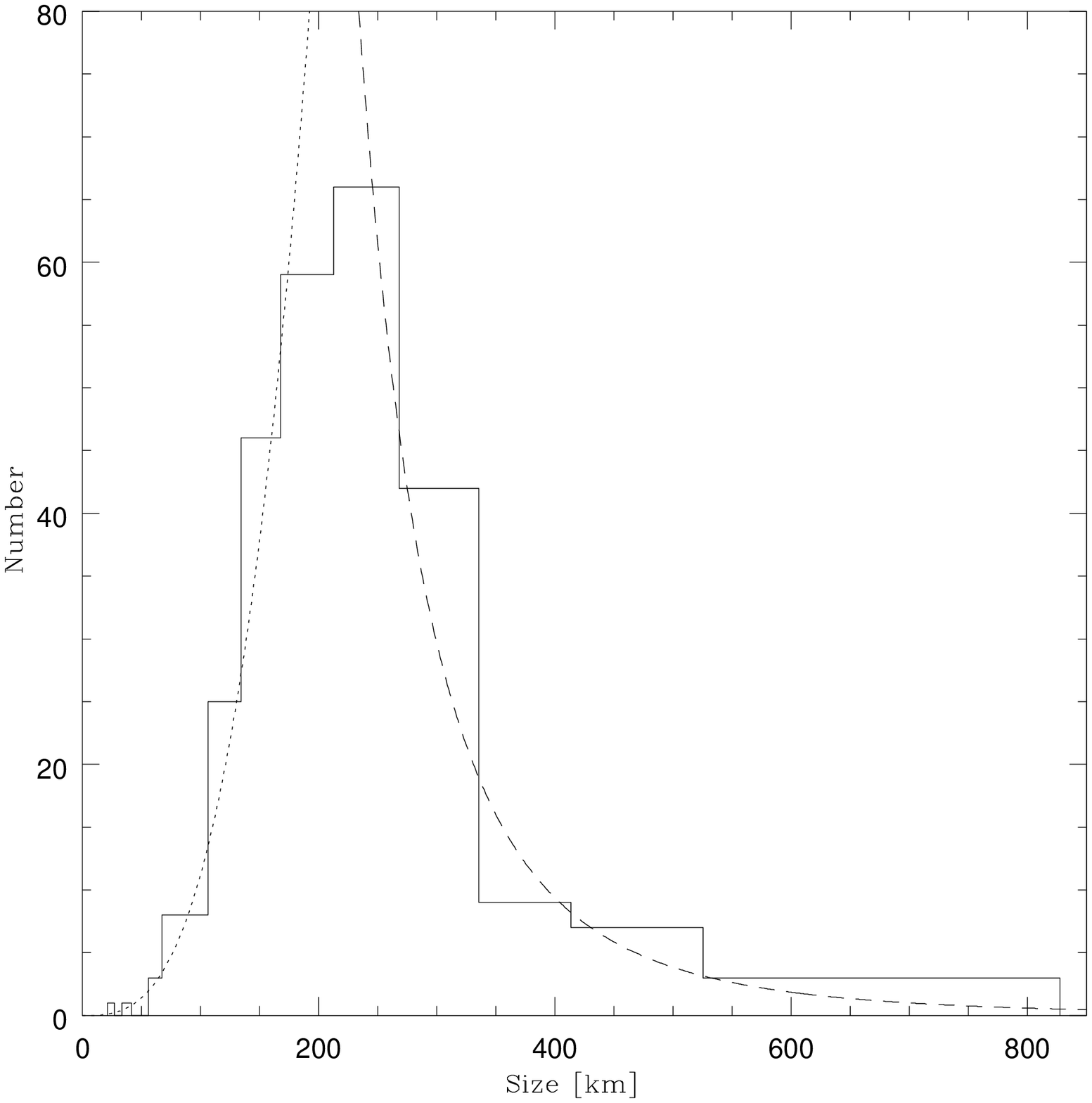}
\caption{The histogram of number of objects against their diameters 
for all current known KBOs, where the dashed and dotted 
lines are the fitting curves (See Equation 1 \& 2).}
\end{figure}

\begin{figure}[tbhp]
\epsfysize 7.0in \epsffile{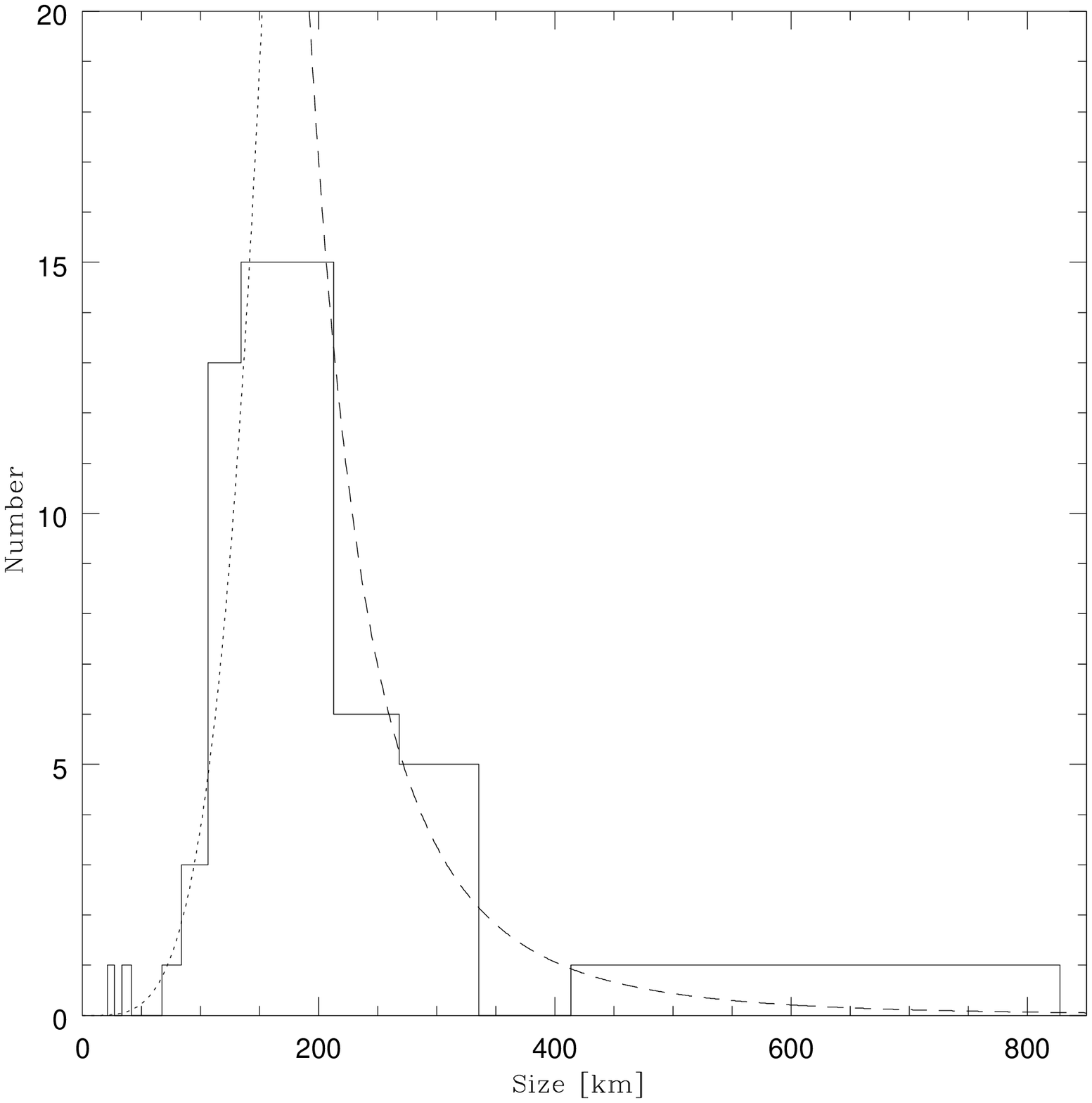}
\caption{The histogram of number of objects against their diameters
for all current known Plutinos, where the dashed and dotted 
lines are the fitting curves (See Equation 3 \& 4).}
\end{figure}

\begin{figure}[tbhp]
\epsfysize 7.0in \epsffile{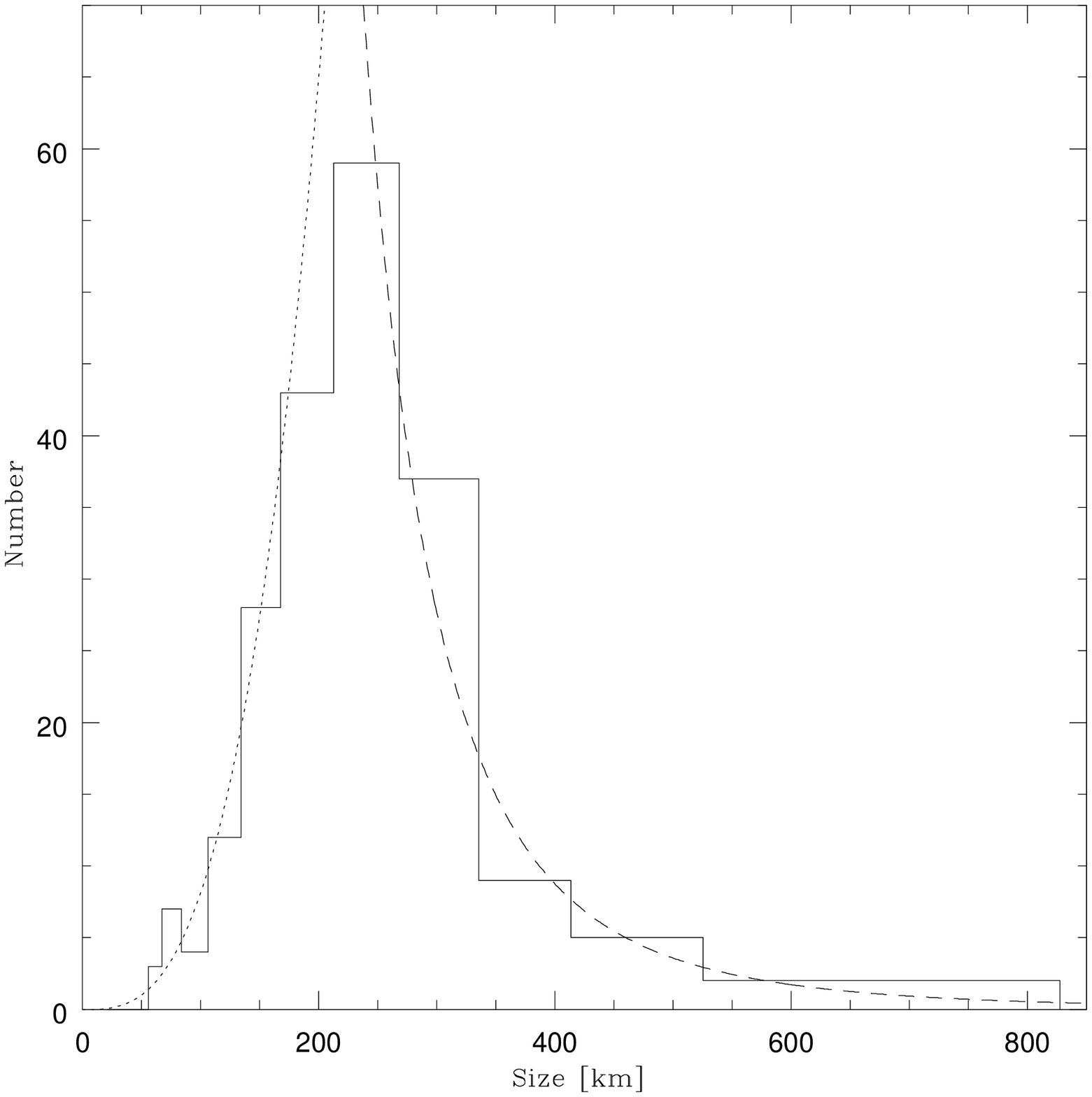}
\caption{The histogram of number of objects against their diameters
for all current known Classical KBOs, where the dashed and dotted 
lines are the fitting curves (See Equation 5 \& 6).}
\end{figure}

\end{document}